\documentstyle[12pt,aasms4]{article}

\begin{document}

\title{The Density and Temperature of Molecular Clouds in M33}
\author{Christine D. Wilson\altaffilmark{1}}
\author{Constance E. Walker\altaffilmark{2}}
\author{Michele D. Thornley\altaffilmark{3}}
\bigskip
\centerline{\it Submitted to the Astrophysical Journal: July 23, 1996}  
\centerline{\it Revised: January 21, 1996} 

\altaffiltext{1}{Department of Physics and Astronomy, McMaster University,
Hamilton, Ontario L8S 4M1 Canada} 
\altaffiltext{2}{Department of Astronomy,  University of Arizona,
Tucson AZ 85721 U.S.A.} 
\altaffiltext{3}{Department of  Astronomy,  University of Maryland,
College Park, MD 20742 U.S.A.} 

\begin{abstract}

We have observed the $^{12}$CO J=2-1, $^{13}$CO
J=2-1, and $^{12}$CO J=3-2 lines in a 
sample of seven giant molecular clouds in
the Local Group spiral galaxy M33 using the
James Clerk Maxwell Telescope. 
The $^{12}$CO/$^{13}$CO J=2-1 line ratio is constant across
the entire sample, while
the observed $^{12}$CO J=3-2/J=2-1 line ratio 
has a weak dependence on the star formation environment of
the cloud, with large changes in the line ratio seen only for clouds
in the immediate vicinity of an extremely luminous HII region.
A large velocity gradient
analysis indicates that clouds without HII regions have
temperatures of 10-20 K,
clouds with HII regions have temperatures of 15-100 K,
and the
cloud in the giant HII region has a temperature of at least
100 K. 
Interestingly,
the giant HII region appears capable of raising the kinetic temperature
of the molecular gas only for clouds that are quite nearby ($< 100$ pc).
The continuous change of
physical conditions across the observed range of star formation environments
 suggests that the unusual physical conditions 
in the cloud in the giant HII region are due to post-star formation
changes in the molecular gas, rather than intrinsic properties of
the gas related to the formation of the giant HII region.
The results from this study of M33 
suggest that similar
observations of ensembles of giant molecular clouds
in more distant normal spiral 
galaxies are likely to give meaningful measurements of the
average physical conditions inside the molecular clouds.
These results also imply that clouds with a factor of
three difference in metallicity have similar density and temperature,
which in turn imply that the 
differences in the CO-to-H$_2$ conversion factor seen in these
clouds can be attributed entirely to metallicity effects.

\end{abstract}


\section{Introduction}

Determining the physical conditions inside molecular clouds is 
important for understanding the link between the properties of
the molecular gas and the types and amount of stars that are formed.
Cloud properties  that could affect the star
formation process include the temperature and 
density of the molecular gas, as well as
the mass fraction in high density gas.
For example, higher
gas temperatures might be required to form high-mass stars
(\markcite{t84}Turner 1984), while 
 a cloud with a  higher mass
fraction of dense gas might form stars with a higher efficiency
(e.g. \markcite{L91}Lada et al. 1991).
In return, star formation, particularly
massive star formation, can affect conditions
inside molecular clouds by compressing the gas
at the boundaries of stellar wind or supernova shocks and
heating the gas by increasing the ultraviolet radiation field.
Giant HII regions are particularly 
interesting targets because they concentrate many hundreds of OB stars
in a relatively small volume of space and represent the only
nearby prototypes for the extreme star formation conditions in
starburst and ultraluminous galaxies. 

In our own Galaxy, observations of low-luminosity HII regions suggest that 
single high-mass stars can form from relatively low-mass molecular
clouds (\markcite{hm90}Hunter \& Massey 1990).
Observations of several nearby giant 
molecular clouds show significant differences in
the spatial distribution of star formation. In the Taurus-Auriga
region, star formation
occurs throughout the cloud as single stars or in small clusters
(\markcite{k90}Kenyon et al. 1990), while Orion B has
most of its star formation occurring in
four massive dense cores (\markcite{L91}Lada et al. 1991), and
Orion A and Ophiuchus are each forming most of their stars
in a single dense cluster (\markcite{ms94}McCaughrean \& Stauffer 1994, 
\markcite{wly89}Wilking, Lada, \& Young 1989).
Despite the differences in the spatial distribution 
of the young stars, the star formation efficiency is 1-4\%
for all four regions (\markcite{el91}Evans \& Lada 1991).
Interestingly, the star formation efficiency in two 
giant HII regions in M33 is also in the range of 2-5\%
(\markcite{wm95}Wilson \& Matthews 1995).

The clumpy structure of molecular clouds (e.g. \markcite{SG90}Stutzki
\& G\"usten 1990) implies that the volume-averaged density of a cloud
is not a good measure of the density of the molecular gas
producing most of the observed emission. For extragalactic
molecular clouds, their low and usually unknown filling factor within
the beam also implies that the kinetic temperature cannot be estimated from the
peak brightness temperature of the observed line. Instead, the density and
temperature must be determined using radiative transfer models and observations
of many line ratios 
of CO and its isotopomers. Because this technique requires observations
of some of 
the rarer isotopomers such as $^{13}$CO, it was first 
applied to starburst galaxies with strong CO lines
(e.g. \markcite{t91}Tilanus et al. 1991,
\markcite{G93}G\"usten et al. 1993). However, one difficulty with
even the nearest starburst galaxies is that the observations measure
the emission from many molecular clouds within a single beam. For
example, at a distance of 3 Mpc, a
20\arcsec~ beam subtends a diameter of 290 pc, which is 
much larger than the 10-100 pc diameter of giant molecular clouds 
in the Milky Way (\markcite{sss85}Sanders, Scoville, \& Solomon 1985).
If the physical conditions in molecular clouds vary significantly from
one cloud to the next, the results of fitting the average line emission
may bear little resemblance to even the average properties of the molecular
cloud ensemble.

Local Group galaxies offer an advantage for such studies
because they are close enough (50 kpc - 1 Mpc) that it is
possible to isolate {\it individual} 
molecular clouds within a 15-20\arcsec~ beam.
In addition, individual  clouds can cover a substantial fraction of the beam
and thus the CO lines  are reasonably strong.
Observations of a large sample of individual clouds can identify
variations in the density and temperature from one cloud to
the next, as well as correlations in cloud properties with the
presence and intensity of massive star formation. The uniformity
(or lack thereof) of the molecular cloud population  also provides a means to
test the reliability of radiative transfer techniques for determining
density and temperature in more distant galaxies, where only
the average emission  from the molecular cloud population can be observed.
In addition, 
density and temperature measurements for individual clouds
can  be used to assess whether changes in the density and brightness 
temperature of the gas are likely to produce systematic errors 
in the observed correlation of the CO-to-H$_2$ conversion factor
with metallicity (\markcite{w95}Wilson 1995, \markcite{dss86}Dickman,
Snell, \& Schloerb 1986, \markcite{s96}Sakamoto 1996). 

In this paper we present observations of
seven giant molecular clouds in the spiral galaxy M33. The clouds were 
chosen to cover a wide variety
of star formation conditions, from clouds with no optical HII regions
to a cloud located in the brightest giant HII region in the galaxy.
The clouds also cover a range of three in oxygen abundance
(\markcite{v88}Vilchez et al. 1988). The observations
and data reduction are discussed in \S 2. The observed line ratios are
compared with previous Galactic and extragalactic observations and
with the star formation environment of the clouds in
\S 3. The density and temperature obtained from an analysis of
the line ratios with a large velocity gradient code are presented
in \S 4. The implications of the results for observations of
more distant galaxies and the calibration of the CO-to-H$_2$ conversion
factor as a function of metallicity are discussed in \S 5. The paper
is summarized in \S 6.

\section{Observations and Data Reduction}

Seven giant molecular clouds in M33 were observed with
the James Clerk Maxwell Telescope (JCMT) 
in the $^{12}$CO J=2-1, $^{13}$CO J=2-1, and $^{13}$CO J=3-2
rotational transitions over four separate
observing runs (1993 July 30 - August 1; 1993 August 25-26; 1994  July 16-18;
and 1995 July 31 -  August 1). 
The half-power beamwidth
of the JCMT is 22\arcsec~ at 230 GHz and 15\arcsec~ at 345 GHz.
The clouds were
selected from the interferometric samples of \markcite{ws90}Wilson \&
Scoville (1990, 1992) and have also been observed in the
$^{12}$CO and $^{13}$CO J=1-0 lines with 55\arcsec~ resolution 
(\markcite{WW94}Wilson \& Walker 1994). 
One position (MC 1) includes emission from two other clouds near
the half-power point of the 22\arcsec~ beam (MC 2, MC 4, \markcite{ws90}
Wilson \& Scoville 1990), while 
only a single cloud produces the CO emission at the other positions.
Each cloud was observed at a single position in the $^{12}$CO J=2-1 
and $^{13}$CO J=2-1 lines and in a five-point cross spaced by 
8\arcsec~ in the $^{12}$CO J=3-2 line. All observations were obtained
 by position-switching to a location in the inner
galaxy ($\alpha(1950)$=01:31:16.9, $\delta(1950)$=30:24:54) 
that had been determined previously to be
 free of $^{12}$CO J=1-0 and J=2-1 emission. Observations of
the $^{13}$CO J=2-1 line were obtained at two or more  different
velocities per source 
to avoid the problem of spikes in the spectrometer masquerading
as a weak line. 
Except for the observations in 1993 July, which 
were obtained with the Canadian Acousto-Optical Spectrometer, all
observations were obtained using the Dutch Autocorrelation Spectrometer.
A log of the observations is given in Table~\ref{tbl-1}.

The calibration  was monitored by  frequently observing
 both planets and spectral line calibrators. 
The main beam efficiencies determined by observing Saturn,
Mars, and Jupiter during the runs in 1993 and 1995
agreed well with the published main beam
efficiencies, and thus for these runs we adopt
the main beam efficiencies from the JCMT User's Guide of 0.69 at 
220 and 230 GHz and
0.58 at 345 GHz. The
spectral line calibrators  were the
evolved stars CRL 2688, CRL 618, and the compact HII region
NGC 7538 IRS1. 
The calibrator observations
agree very well with reference spectra available at the JCMT, with
an average 
ratio of the observed  to reference peak temperature of 1.00 with an
rms scatter of 13\%.
The rms dispersion of the individual measurements is 10\% for
the $^{12}$CO observations and 20\% for the $^{13}$CO observations.
We attribute the larger scatter in the $^{13}$CO observations
to the weakness of the lines and adopt 10\% as our absolute
calibration uncertainty in any single line on the basis of
the $^{12}$CO observations.

Throughout 
the 1994 run, the peak temperatures observed for CRL 618 and CRL 2688
were consistently 70\% of the values measured in the other runs. The
low efficiency of the telescope was confirmed by remeasuring the 
$^{12}$CO J=2-1 spectrum of MC 20, which was also 70\% of the previously
measured value.
The main beam efficiency determined from observations
of Mars and Uranus was also low by the same amount.
Thus for the data obtained during this run, we
adopt a main beam efficiency at 220 and 230 GHz of $0.70 \times 0.69 = 0.48$.

As we are comparing line strengths measured with the same beam diameter
(either the original beam or convolved to a larger beam), the best
temperature scale to use would be the $T_R^*$ scale. Although the
clouds in M33 are much smaller than the full diffraction pattern of
the telescope (5-16\arcsec), the $T_R^*$ scale corrects for
all telescope efficiencies except the source-beam coupling efficiency,
which should be the same at all frequencies since we have observed
the emission with the same beam at each frequency. However, to
convert the observed temperatures $T_A^*$ to $T_R^*$ requires
knowing $\eta_{fss}$, the forward scattering and spillover efficiency.
This efficiency is difficult to measure, and so
we did not attempt to measure $\eta_{fss}$ during our observing runs. As a
result, we do not know how to correct the published values of $\eta_{fss}$
for the calibration problems encountered in the 1994 run. However, we do have
good measurements of the main beam efficiency for each run and
under normal calibration conditions, adopting the main beam temperature scale 
would only change the $^{12}$CO J=3-2/J=2-1 line ratios by 5\% compared
to the $T_R^*$ scale. Thus we use the main beam
temperature scale throughout this paper, and in particular 
in calculating the line ratios for comparison
with radiative transfer models.

Given the angular size of the M33 molecular clouds  (5-16\arcsec),
the main beam temperature, $T_{MB}$,
is a better approximation to the brightness or radiation 
temperature, $T_R$, than is the antenna temperature, $T_A^*$. The
three temperature scales are related by
$$T_R = { T_A^* \over \eta_c \eta_{fss} } = { T_{MB}
\eta_{MB} \over \eta_c \eta_{fss} } $$
where $\eta_c$ is the source-beam coupling factor, $\eta_{fss}$ is the
forward scattering and spillover efficiency, and $\eta_{MB}$ is
the main beam efficiency. We can estimate $\eta_c$ for the M33 clouds
since they have been observed at high resolution (\markcite{ws90}Wilson 
\& Scoville 
1990, 1992). For the 230 GHz observations, $\eta_c > 0.9$, while
$\eta_{fss} = 0.80$ and $\eta_{MB} = 0.69$, so that
$T_R \sim 0.9T_{MB}-1.0T_{MB}$.  At 345 GHz, $\eta_c > 0.8$, while
$\eta_{fss} = 0.70$ and $\eta_{MB} = 0.58$, so that
$T_R \sim 0.8T_{MB}-1.0T_{MB}$.  We use the main beam
temperature scale throughout this paper, and in particular 
in calculating the line ratios for comparison
with radiative transfer models.

The data were reduced using the Bell Labs data reduction package
COMB. The data were binned to a resolution of 1 km s$^{-1}$ and
first to third order baselines removed. The lines are quite narrow
($\sim 20$ km s$^{-1}$) and the spectrometer bandwidth was always
at least 200 km s$^{-1}$, so the  higher order
baselines  should not introduce significant errors into the final
line intensities. Each set of five
 $^{12}$CO J=3-2 spectra were then convolved to simulate
a 22\arcsec~ beam. The spectra for
each cloud are shown in Figure~\ref{fig-1}. 
Channel-by-channel ratios of the two spectra were
used to measure the average line ratio for each cloud. Only channels with
a signal-to-noise ratio of at least 2 in the line ratio were used.
This restriction resulted in
 the $^{12}$CO/$^{13}$CO J=2-1 line ratio being obtained from the
central 5-14 km s$^{-1}$ ($0.6-1.3 \times \Delta V_{FWHM}$), while
the $^{12}$CO J=3-2/J=2-1 line ratio was obtained from the central
9-19 km s$^{-1}$ ($1.1-1.7 \times \Delta V_{FWHM}$).
The measured line ratios 
 and  $^{12}$CO J=2-1 integrated intensity for each cloud
are given in Table~\ref{tbl-3}. 

\section{CO Line Ratios and the Star Formation Environment}

The $^{12}$CO/$^{13}$CO J=2-1 line ratios given in Table~\ref{tbl-3} 
have an average value of 7.3 and an rms dispersion
of 1.3 (18\%). 
Thus we conclude that there is
no intrinsic variation in the $^{12}$CO/$^{13}$CO J=2-1 line ratio
from one cloud to another in M33. This average line ratio compares quite
well with  values measured in other  galaxies
such as M51 (6-10 in center and spiral
arms, \markcite{gb93}Garcia-Burillo, Guelin, \& Cernicharo
 1993) and NGC 4414 ($9\pm 1$,
\markcite{b93}Braine, Combes, \& van Driel 1993). The value in M33
also agrees quite well with the value  measured in a pencil-beam
survey of the Milky Way ($5.5\pm 1$,\markcite{s93}Sanders et al. 1993) and
in large-area maps of the Orion A and B molecular clouds (4-6 along the
main ridge, \markcite{s94}Sakamoto et al. 1994). The M33 value is
slightly smaller than the value measured in the starburst galaxy
M82 ($12\pm 3$, \markcite{t91}Tilanus et al. 1991) and in
a large sample of starburst and interacting galaxies
($13\pm 5$, \markcite{a95}Aalto et al. 1995). Previous measurements
have indicated that the $^{12}$CO/$^{13}$CO J=1-0 line ratio is also
somewhat larger in starburst and
merging galaxies compared with other galaxies
(\markcite{bf91}Becker \& Freudling 1991; 
\markcite{a91}Aalto et al. 1991; \markcite{c91}Combes et al. 1991;
\markcite{ca91}Casoli et al. 1991).

The $^{12}$CO J=3-2/J=2-1 line ratios given 
in Table~\ref{tbl-3} show more scatter.
In particular, the line ratio obtained for NGC 604-2 ($1.07\pm0.03$ 
is significantly
higher than the value for any other cloud. Excluding NGC 604-2, the
average $^{12}$CO J=3-2/J=2-1 line ratio is 0.69 with an rms
dispersion of 0.15 (21\%). 
This average line ratio for the six clouds in M33
is in good agreement with values  measured in
other  galaxies such as the starburst galaxies M82 ($0.8\pm0.2$, 
\markcite{g93}G\"usten et al. 1993),
NGC 253 ($0.5\pm0.1$ in the disk, \markcite{w91}Wall et al. 1991),
and IC 342 ($0.47\pm0.12$, \markcite{ia93}Irwin \& Avery 1993). 
The average value for the six clouds in M33 
also agrees well with values measured in M51 (0.7 in the spiral
arms, \markcite{gb93}Garcia-Burillo et al. 1993) and in a  
 pencil-beam
survey of the Milky Way ($0.55\pm0.08$, \markcite{s93}Sanders et al. 1993).
It also agrees with
a previous measurement for M33 in a slightly larger beam 
($0.64\pm0.28$, \markcite{tw94}Thornley \& Wilson 1994).
Higher line ratios similar to the value for the cloud NGC 604-2
are seen in the nuclei of both NGC 253 ($\sim 1$, 
\markcite{w91}Wall et al. 1991)
and M51 (1.1, \markcite{gb93}Garcia-Burillo et al. 1993). 

The clouds observed in M33 have oxygen abundances that range
over a factor of three (\markcite{v88}Vilchez et al. 1988). If we
momentarily 
ignore the unusually high $^{12}$CO J=3-2/2-1 line ratio in NGC 604-2,
 there is no evidence that the observed
line ratios depend on the metallicity of the cloud. In particular,
the two line ratios for MC 19 and NGC 604-4 are very similar despite
a factor of three difference in the oxygen abundance
(Table~\ref{tbl-3}). NGC 604-2 and
NGC 604-4 are located within 120 pc of each other and presumably share
the same oxygen abundance, so we attribute the 
difference in the $^{12}$CO J=3-2/2-1 line ratios for these two
clouds to some factor other than metallicity, most likely the star
formation environment. 

If we examine the star formation environments of our cloud sample, 
two of the clouds (MC 19 and MC 32) lie
between 100 and 150 pc away from the nearest optical HII region 
(\markcite{ws91}Wilson \& Scoville 1991), while
 three clouds (MC 1, MC 13, and MC 20)
contain optical HII regions. The cloud NGC 604-2
is located near the center of the giant HII region NGC 604,
on the southern edge of the OB association which presumably powers
the HII region (\markcite{wm95}Wilson \& Matthews 1995), 
while the cloud NGC 604-4 is located at the very edge of the southern
extent of the H$\alpha$ emission from the HII region. 
If we compare the CO line ratios
for the three clouds without HII regions (MC 19, MC 32, and
NGC 604-4) with those for the clouds
with normal HII regions (MC 1, MC 13, and MC 20), the
average $^{12}$CO/$^{13}$CO J=2-1 line ratios for the two
sets of clouds agree very well. 
The $^{12}$CO J=3-2/J=2-1 line
ratios for the clouds with optical HII regions ($0.79\pm0.05$) are somewhat
higher  than those for the clouds without optical 
HII regions ($0.58\pm0.06$), while the line ratio for NGC 604-2 in the giant
HII region is even higher ($1.07\pm0.03$). 
The difference between the two clouds (NGC 604-2 and NGC 604-4)
in and around the giant
HII region is striking. The projected separation of the two clouds 
is only 120 pc (30\arcsec), yet only the cloud in the
 immediate vicinity  of the giant HII region and its powering OB
association has an unusual line ratio.

The high $^{12}$CO J=3-2/J=2-1 
line ratio in NGC 604-2 may be a clue to 
 unusual physical conditions in the molecular clouds from which
the HII region NGC 604 formed,
 i.e.  a pre-star formation difference in the physical conditions that may have
 caused the formation of the giant HII region.
Alternatively,
 the high line ratio may be due to heating of the gas by the massive
 stars, i.e. a post-star formation change in the physical conditions in
 the molecular cloud. This interpretation implies that
 the giant HII region has a relatively
 small sphere of influence over which its intense radiation field can
 change the properties of the dense molecular gas. 
 It appears that both an unusually intense radiation field and a cloud
 in close proximity to the source of the ionizing radiation are required
 to produce a large change in the CO line ratios.
However, the $^{12}$CO J=3-2/J=2-1 
line ratios observed for the clouds with optical HII regions 
are higher than those for clouds without HII regions at the
$2\sigma$ level. This difference suggests that
HII regions that are more than an order of magnitude less luminous than 
NGC 604 can produce smaller changes in  the CO line ratios. Thus the
data for M33 provide some evidence for a continuous change of line 
ratio with increasing intensity in the star formation environment.

One complication to this interpretation is that the optical HII
regions could in principle be separated from the molecular
cloud by up to the scale height of the disk ($\sim 200$ pc) and
simply appear projected on the cloud. However, the fact that most
HII regions  appear close to molecular clouds
(\markcite{ws91}Wilson \& Scoville 1991)
suggests that there is a true physical association between the clouds and
the HII regions. Also,
the three clouds without optical HII regions could actually 
be forming stars, since
the clouds have sufficient optical depths to hide an HII region if it
were located on the far side of the cloud. However, since the $^{12}$CO lines
are optically thick for these clouds, we can only see emission from
the near and presumably cool
side of the cloud and so whether or not the clouds have an HII
region on their far side is irrelevant to this analysis. The increasing
line ratio with increasing HII region luminosity is consistent with
a picture in which  the ultraviolet radiation field of the optically
visible HII region heats the molecular gas.
We conclude that the $^{12}$CO J=3-2/J=2-1 line ratio of a cloud
has a weak dependence on the star formation environment of
the cloud, with large changes in the line ratio seen only for clouds
in the immediate vicinity of an extremely luminous HII region.

\section{Density and Temperature Determined from Large Velocity Gradient
 Models}

\subsection{The Large Velocity Gradient Model}

There are several different assumptions that can be made  in
calculating source parameters from molecular line data. The most
common assumption is that the gas is in local thermodynamic 
equilibrium (LTE), where the number density of molecules in any
particular energy level is only a function of temperature. 
Probably the next most common assumption is that there is
 a large velocity gradient (LVG) in the gas cloud
(e.g. \markcite{ss74}Scoville \& Solomon 1974). The presence of 
such a gradient has the effect of Doppler-shifting the emission of a
molecule relative to that of molecules elsewhere in the cloud. 
When the mean free path of a line photon exceeds the local velocity 
dispersion divided by the velocity gradient, it can travel through
the cloud without being absorbed elsewhere.  A third approach is to
 assume that the velocity field is microturbulent. In
microturbulent radiative transfer models, the 
velocity structure is due to thermal motions and small-scale
random motions (\markcite{w77}White 1977). Unlike the LVG case,
neighboring cloud components are radiatively coupled, with the degree to
which they are coupled depending on the magnitude of the random
motions and the amount of resonance scattering. 
A fourth possibility is to use 
 a photon mean escape probability formalism 
to simplify radiative transfer calculations. In this approach,
level populations and the internal intensity are calculated for
an ``average'' location in a uniform, spherical cloud, with the photon
escape probability from this position being determined 
by the optical depth (Black \& 
Aalto 1991, private communication). These latter three non-LTE techniques
often yield similar results when applied to galactic molecular
clouds (\markcite{w77}White 1977) and thus we have chosen to use the
LVG formalism in this analysis. 

To estimate the physical conditions in the molecular clouds in M33, 
we used the large velocity gradient code RAD written by Lee Mundy and
implemented as part of the MIRIAD data reduction package. Models were
run for kinetic temperatures $T_K = 10,15,20,30,50,100,200,300$ K
and for three different values of the [$^{12}$CO]/[$^{13}$CO] abundance
ratio (30, 50, 70) to span the range of abundances seen in
the Milky Way (\markcite{lp90}Langer \& Penzias 1990).
For each kinetic temperature and abundance combination, the 
H$_2$ density range was
$n_{H_2} = 10-10^6$ cm$^{-3}$ 
and the $^{12}$CO column density range per unit velocity was
$N(^{12}\rm CO)/\Delta V = 10^{15}-10^{20}$ cm$^{-2}$ (km s$^{-1}$)$^{-1}$.
We use the full-width half maximum velocity to calculate the 
$^{12}$CO column density from $N/\Delta V$, where the average
velocity width for the six clouds excluding NGC 604-2 is
$9.3\pm0.6$ km s$^{-1}$.

One drawback of both the LTE and non-LTE techniques described here
is that they do not take into account the effects of cloud-cloud shielding. 
Cloud-cloud shielding occurs when two or more molecular clouds are aligned 
along our line of sight in both space and velocity. In this case the
radiation from the background cloud(s) may be fully or partially
absorbed by the foreground cloud. When a pair of isotopes, such as
$^{12}$CO and $^{13}$CO, are being used to probe the optical depth along a 
line of sight through a galaxy, cloud-cloud shielding will tend to produce
{\it overestimates} of the optical depth in individual clouds. 
Optical depth estimates through the ensemble of clouds are
only affected by cloud-cloud shielding when
the rarer isotope ({\it e.g.} $^{13}$CO) becomes optically
thick and so does not sample all the clouds along the line of sight. 
Cloud-cloud
shielding is unlikely to be a significant 
problem in M33, which has a relatively
low molecular gas surface density and is not highly inclined.

\subsection{Model Results for M33 Molecular Clouds}

To provide better constraints for the models, we include the
$^{12}$CO/$^{13}$CO J=1-0 line ratio observed for these clouds in
a 55\arcsec~ beam (\markcite{ww94}Wilson \& Walker 1994). 
(The possibility that the larger beam size of the J=1-0 data can produce
systematic errors in our results is discussed later in this section.)
Although
both the NGC 604 clouds were observed within a single beam, we can
use their significantly different central velocities to isolate the
line ratio for each cloud.
The average $^{12}$CO/$^{13}$CO J=1-0 line ratio for the seven
clouds is $9.7\pm0.6$ with an rms dispersion of 1.5 or 15\%.
We combine this average J=1-0 line ratio with the average J=2-1 line
ratio in our modeling. Since there appears to be a trend in 
 the $^{12}$CO J=3-2/J=2-1 line ratio with star formation environment,
we used four values of this line ratio in our analysis: the average
value for all six clouds ($0.69\pm0.06$); the average value for
clouds without HII regions ($0.58\pm0.06$); 
the average value for clouds with HII regions ($0.79\pm0.05$);
and the value for NGC 604-2 ($1.07\pm0.16$),
where the total uncertainty in the line ratio for NGC 604-2 has been estimated
at 15\% to match the expected calibration uncertainty in the line ratio.

Acceptable solutions for each  set of three line ratios 
($^{12}$CO/$^{13}$CO J=1-0, $^{12}$CO/$^{13}$CO J=2-1, and
$^{12}$CO J=3-2/J=2-1) 
were found
by visual inspection of plots of the line ratios as a function of density
and column density at each kinetic temperature. 
Acceptable solutions were defined to be ones for
which all three line ratios agreed 
within their $1\sigma$ uncertainties. Examples of representative
solutions are given in Figure~\ref{fig-2} and the full range of solutions
is given in Table~\ref{tbl-4}. The ranges of values given for
temperature, density, and column density 
for each [$^{12}$CO]/[$^{13}$CO] abundance ratio are correlated in the
sense that solutions with higher temperatures have higher column densities
and lower densities.

The results from the LVG analysis show that 
the kinetic temperature increases from clouds without HII
regions ($T_K = 10-20$ K) to clouds with HII regions
($T_K = 15-100$ K) to NGC 604-2 ($T_K \ge 100$ K).
The column density also increases by
at least an order of magnitude from clouds without HII regions
to NGC 604-2. The molecular gas in all seven clouds
is quite dense ($10^3-3\times 10^4$ cm$^{-3}$), although the
density is lower in NGC 604-2 than in the clouds without HII regions.
Thus an increase in the 
 $^{12}$CO J=3-2/J=2-1 line ratio  produces LVG solutions 
 with higher kinetic temperatures, higher column densities, 
and lower densities. 
The volume-averaged densities (mass divided by volume) 
are 40-210 cm$^{-3}$ for the three clouds without HII regions
and $\sim 200-400$ cm$^{-3}$ for the three clouds with HII regions
(calculated from
\markcite{ws90}Wilson \& Scoville 1990, 1992), 
significantly lower than the densities derived
from the LVG analysis ($5\times 10^3-3\times 10^4$ cm$^{-3}$
and $2\times 10^3 - 10^4$ cm$^{-3}$, respectively).
The volume filling factor of the dense
gas is $0.1-4\%$ for clouds without HII regions and $\sim 2-20\%$
for clouds with HII regions.
For NGC 604-2 the volume-averaged density is somewhat higher,
$\sim 500$ cm$^{-3}$, which combined with the somewhat lower 
density from the LVG model of $10^3-3\times 10^3$ cm$^{-3}$ 
gives a filling factor for the dense gas of 17-50\%.
These results compare well with the volume filling factor of
10\% for clumps in Orion (\markcite{s94}Sakamoto et al. 1994).

Are the unusual physical conditions in NGC 604-2 due to the presence
of the giant HII region, or do they represent pristine conditions
in the molecular gas that could have played a role in initiating the
formation of the giant HII regions?
An increase in the kinetic temperature of the gas could be easily 
accomplished by the intense ultraviolet radiation field
produced by the dense cluster of OB stars powering the giant HII region.
The gas column density could be
increased through shock-initiated merging of two or more molecular
clouds or through a partial collapse of the molecular cloud due to
an increase in the external gas pressure. However, the molecular
column density could also be decreased through photo-dissociation of the 
molecular gas and thus it is difficult to predict the net effect
on the column density from the formation of the giant HII region.  
It is difficult to understand how the formation of the giant HII region could
have produced the higher filling factor and lower density of the
dense clumps in the molecular cloud. One possible scenario would
be that the formation of the dense star cluster depleted the gas in
the densest clumps, leaving behind the lower density
inter-clump gas. This gas could then be compressed to somewhat higher density
as the cloud shrank due to the increased external gas pressure.
In any case, the continuous range  of {\it all} physical conditions
(kinetic temperature, column density, density, and filling factor)
observed from clouds without HII regions to clouds with normal
HII regions to this cloud in a giant HII
region suggests that the different physical conditions in NGC 604-2
are due to post-star formation changes in the molecular gas, rather
than intrinsic conditions of the pre-star formation molecular cloud.

The similarity
of the CO line ratios for the six clouds in the normal disk 
despite a range of a factor of
three in oxygen abundance might naively suggest
that the [$^{12}$CO]/[$^{13}$CO] abundance ratio does not change
significantly when the oxygen abundance is varied by a factor of three. 
However, this conclusion is in contrast
to the Milky Way, where \markcite{lp90}Langer \& Penzias (1990) have
measured a significant radial gradient in the [$^{12}$CO]/[$^{13}$CO] 
abundance ratio. In fact, for kinetic temperatures of 10-20 K there
do exist possible LVG solutions with different values for the
[$^{12}$CO]/[$^{13}$CO] abundance ratio but with the same density,
kinetic temperature, and $^{13}$CO column density which produce
identical CO line ratios (Figure~\ref{fig-3}). This result can be
easily understood as being due to the low optical depth of the
$^{13}$CO emission and the high optical depth of the $^{12}$CO emission.
As long as the $^{13}$CO column density is fixed, the 
observed line ratios will be the same for a large range in the
[$^{12}$CO]/[$^{13}$CO] abundance ratio. Thus our data are consistent
with M33 containing a similar radial gradient in the
[$^{12}$CO]/[$^{13}$CO] abundance ratio as found in
the Milky Way, without requiring any
conspiracy between several physical parameters to produce the observed
constant CO line ratios. 

Using the average $^{12}$CO J=3-2/J=2-1
line ratio for the six clouds (excluding NGC 604-2) in
the LVG analysis produces solutions
for temperature, density, and 
column density that are intermediate between those obtained
for the clouds with and without HII regions separately (Table~\ref{tbl-4}). 
The mean value for each parameter derived using the average line
ratio is within a factor of two of the average of the mean parameters
derived for clouds with and without HII regions separately.
Thus physical conditions derived by averaging together emission from
clouds both with and without HII regions provide a reasonable approximation
to the average physical conditions in this group of six molecular clouds.

\subsection{Possible Sources of Systematic Errors}

There are several possible complications that could introduce
systematic errors into the LVG results. For example, 
the difference in the beam area used in the 
$^{12}$CO/$^{13}$CO J=1-0 ratio compared
with the other ratios could introduce systematic errors 
 if the J=1-0 line ratio varies as a function of the 
spatial resolution used (\markcite{ww94}Wilson \& Walker 1994). We can assess
what effect this would have by looking at the LVG results
using  a lower value 
of $5\pm 1$, which is   similar to the values of 3 to 6
seen in Galactic molecular clouds
(\markcite{gb76}Gordon \& Burton 1976, \markcite{sss79}Solomon, Scoville,
\& Sanders 1979). 
For simplicity we will use the results for the average of all six clouds
in this analysis.
For a [$^{12}$CO]/[$^{13}$CO] abundance ratio of 50 and 
 this lower J=1-0 line ratio, the solutions permitted for the six
clouds are $T_K = 20-300$ K, $n = 60-2000$ cm$^{-3}$, and
$N(^{12}\rm CO) = 9\times 10^{17}-3\times 10^{19}$ cm$^{-2}$.
 The solutions permitted for NGC 604-2 are
$T_K = 50-300$ K, $n = 30-70$ cm$^{-3}$, and
$N(^{12}\rm CO) = 1-2\times 10^{19}$ cm$^{-2}$,
with an additional possible solution with
$T_K = 300$ K, $n = 600$ cm$^{-3}$, and
$N(^{12}\rm CO) = 2\times 10^{19}$ cm$^{-2}$.
Much of the increased range of permitted solutions can be
attributed to the
larger uncertainty in the lower adopted line ratio (20\%) compared
with the observed line ratio (6\%).
On average, the effect of adopting this lower J=1-0 line
ratio is to increase the kinetic temperature range  for which solutions
are available, to decrease the density
by a factor varying from a few to more than an order of magnitude, 
and to increase the column density and its range.

Yet another possible difficulty is that the sizes of the clouds
may be different in the different transitions. In applying the
LVG analysis we have implicitly assumed that the beam filling factor
is the same in each line. Recent observations of
the Orion molecular cloud show that the cloud appears smaller in the
$^{12}$CO J=2-1 line than in the J=1-0 line, while the
$^{13}$CO J=2-1 line may have an even narrower extent (\markcite{s94}Sakamoto
et al. 1994). If the J=2-1 and J=3-2 emission regions have different
angular extents, we would expect the J=3-2 emission to originate
in the smaller region. In this case, the observed 
$^{12}$CO J=3-2/J=2-1 line ratios
 would set only  lower limits on the true line ratios, 
and so the true kinetic
temperatures would be larger than those derived here. 
In this scenario, the large $^{12}$CO J=3-2/J=2-1 line ratio
in NGC 604-2 could be due to a larger ratio of
the J=3-2 to J=2-1 emitting areas compared with the other clouds.
These hypotheses could be tested with an imaging submillimeter interferometer
such as the Smithsonian Astrophysical Observatory's 
Submillimeter Array, which is presently under construction on Mauna Kea.

Another possible problem is that more than one
temperature component or a smooth range of temperatures may be present in
some or all of the clouds in our sample
(for example, see models by \markcite{gsw92}Gierens, Stutzki,
\& Winnewisser 1992). This is particularly true
for NGC 604-2, which, depending on its three-dimensional 
orientation relative to the giant HII region, could be subject to 
intense heating on only one side. 
Some support for a two-component model is given by the
brightness temperatures predicted by models with
$T_K \ge 100$ K. The brightness
temperature predicted for the $^{12}$CO J=2-1 line is $\ge 60$ K,
which when combined  with the observed peak temperature
of 1.19 K gives a filling factor of
only 2\%. This filling factor corresponds to a source diameter
of only 3\arcsec~ in a 22\arcsec~ beam, 
significantly smaller than the 9\arcsec~ diameter
measured in the $^{12}$CO J=1-0 line (\markcite{ws92}Wilson \&
Scoville 1992). 
A crude two-component model for NGC 604-2 could consist of
one component with line ratios like the six
other clouds and a hot component with
$^{12}$CO J=3-2/2-1 $\sim$2.  This model predicts that 
the two components contribute about equally to
the $^{12}$CO J=3-2 emission,
 while roughly three-quarters of the $^{12}$CO J=2-1
emission arises in the cooler component.
This cool component gives a brightness temperature
of $\sim 5$ K for $T_K = 10$ K, which corresponds to a source
diameter of $\sim 9\arcsec$.
Additional circumstantial evidence in support of more than one
temperature component comes from the measurement of the
$^{12}$CO J=2-1/1-0 line ratio in MC 32 and MC 19 
(\markcite{tw94}Thornley \& Wilson 1994). Despite the 
relatively large measurement uncertainty 
($0.67\pm0.19$ in a 54\arcsec~ beam), this line ratio only overlaps 
the solutions for the  three line ratios discussed here for 
 $T_K = 15$ K and
[$^{12}$CO]/[$^{13}$CO] abundance ratios of 50-70. A more precise 
measurement of this line ratio might easily remove this area of overlap,
which would again suggest that the lowest-J rotational transitions
originate in a cooler component than the J=3-2 transition.

\section{Implications for Distant Galaxies and the CO-to-H$_2$ Conversion
Factor}

The relative 
uniformity of the line ratios of the six molecular clouds observed in
the normal disk of M33 suggests that 
similar observations of ensembles of molecular clouds in more
distant galaxies are likely to produce meaningful measurements of
the average physical conditions of the molecular gas. This conclusion is
also supported by the reasonable agreement between physical conditions
derived using the average line ratios for the six clouds and the
average of the physical conditions derived separately for clouds with
and without HII regions (\S 3).
The relatively
small sphere of influence ($\sim$100 pc) of the giant HII region NGC 604
suggests that in normal galaxies only the most intense star forming
regions may produce significant changes in the molecular gas. The
change in the line ratios is likely to be measurable 
 only in relatively nearby  galaxies ($< 10$ Mpc), where the warm
molecular gas in the HII region is not diluted by emission from
cooler gas included in the beam. It would be interesting to test
whether similarly uniform line ratios are observed in individual clouds
in the intense ultraviolet field of a starburst galaxy, but such
observations must await the construction of an imaging submillimeter
interferometer.

The CO-to-H$_2$ conversion factor, $\alpha$, is predicted to depend on the
density and brightness temperature of the molecular
clouds as $\sqrt{n}/T_B$, as well as on the metallicity 
(\markcite{dss86}Dickman et al. 1986, \markcite{s96}Sakamoto
1996). Observations of Local Group galaxies, including several of
these clouds in M33, have  been used to calibrate the dependence
of the conversion factor on metallicity (\markcite{w95}Wilson 1995).
However, given the generally lower shielding by dust and 
 higher ultraviolet radiation field in low-metallicity
irregular galaxies, it is possible that molecular clouds in lower metallicity 
environments have systematically different brightness temperatures
and densities. We can use our solutions for the physical conditions in
 the molecular clouds in M33 to determine the possible range in 
 $\sqrt{n}/T_B$ and to see whether systematic errors in the metallicity
calibration are likely to exist.

Ignoring the possible effects of 
density and brightness temperature differences,
\markcite{w95}Wilson (1995) obtained
$\alpha \propto (12+log(O/H))^{-0.67}$. 
For the six molecular clouds (excluding NGC 604-2), 
$\sqrt{n}/T_B$ can vary by a factor of six for kinetic
temperature solutions from 10 to 30 K. This large range 
in $\sqrt{n}/T_B$ is produced
by the tendency of the solutions to have lower densities and higher
brightness temperatures for higher kinetic temperatures. As a result, it
is theoretically possible for NGC 604-4, which has a metallicity three
times smaller than the other five clouds, to have $\sqrt{n}/T_B$ 
a factor of six smaller, which would produce a significantly steeper
slope ($\sim -2$) in the derived metallicity calibration.
However, to produce this factor of six difference in $\sqrt{n}/T_B$ 
requires NGC 604-4 to have a [$^{12}$CO]/[$^{13}$CO] abundance ratio
of 30, while the clouds in the inner disk
would need an abundance ratio of 50.
Thus to produce significant systematic errors in the metallicity
calibration of the CO-to-H$_2$ conversion factor would 
require a radial [$^{12}$CO]/[$^{13}$CO] gradient in the opposite
sense from that observed in the Milky Way (\markcite{lp90}Langer \& 
Penzias 1990). In addition, the different densities 
($4\times 10^3$ cm$^{-3}$ vs. $2\times 10^4$ cm$^{-3}$) and excitation
temperatures ( 30 K vs. 10 K) 
for NGC 604-4 and the inner disk clouds would have to
``conspire'' to produce the same observed CO line ratios despite
large physical differences. 
We conclude that the molecular clouds in M33 provide
 no evidence for a systematic error in the metallicity calibration
of the CO-to-H$_2$ conversion factor due to differences in the
density or brightness temperature from low to high metallicity clouds.

\section{Conclusions}

We have observed the $^{12}$CO J=2-1, $^{13}$CO
J=2-1, and $^{12}$CO J=3-2 lines in 
a sample of seven giant molecular clouds in
the Local Group spiral galaxy M33. 
The clouds were chosen to cover a range of star formation conditions,
from clouds without optical HII regions to a cloud in 
a giant HII region. We find that the
$^{12}$CO/$^{13}$CO J=2-1 line ratio is constant across
the entire sample, while the $^{12}$CO J=3-2/J=2-1 line ratio
is somewhat smaller in three clouds without optical HII regions than
in three clouds with HII regions.
The seventh cloud, located in
the brightest giant HII region in the galaxy, has
an even higher $^{12}$CO J=3-2/J=2-1 line ratio.
We conclude that the $^{12}$CO J=3-2/J=2-1 line ratio of a cloud
has a weak dependence on the star formation environment of
the cloud, with large changes in the line ratio seen only for clouds
in the immediate vicinity of an extremely luminous HII region.

We used a large velocity gradient code to determine the density,
temperature, and column density for the clouds. The analysis indicates
that clouds without HII regions have temperatures in the range of 10 to 20 K
and
clouds with HII regions have temperatures in the range of 15 to 100 K,
while the cloud in the giant HII region has a kinetic temperature
of at least 100 K. 
We note that the giant HII region seems to have 
a relatively limited sphere of influence within which it
can heat the molecular gas, since
a molecular cloud located only 120 pc from the giant HII region shows a normal
cooler temperature.
The column density increases by about an order
of magnitude from the cool clouds to the hot cloud, as does the
volume filling factor of the dense gas, while the density
decreases by less than an order of magnitude. The continuity of
physical properties (kinetic temperature, density, column density,
and filling factor) across the range of star formation environments
in our sample suggests that the unusual physical conditions seen
in the cloud in the giant HII region are due to post-star formation
changes in the molecular gas, rather than intrinsic properties of
the gas related to the formation of the giant HII region.

Excluding the giant HII region, the relatively uniform line ratios 
observed in M33 
suggest that average physical conditions determined from similar
measurements of ensembles of giant molecular clouds
in more distant normal spiral galaxies are likely to give physically meaningful
results. These uniform line ratios also imply that the average
brightness temperature and density are similar in clouds ranging over
a factor of three in metallicity, which in turn suggests that
differences in the CO-to-H$_2$ conversion factor seen in these
clouds can be attributed entirely to metallicity effects.

\acknowledgments

The research of CDW is supported through a grant
from the Natural Sciences and Engineering Research Council
of Canada. MDT thanks NRC Canada for travel support. The JCMT 
is operated by the Royal Observatories on behalf of the Particle
Physics and Astronomy Research Council of the United Kingdom, the
Netherlands Organization for Scientific Research, and the National Research
Council of Canada.

\clearpage

\figcaption[fig1.ps]{$^{12}$CO J=2-1, $^{13}$CO J=2-1,
and $^{12}$CO J=3-2 spectra for seven giant molecular clouds
in M33. The spectra are binned to 1 km s$^{-1}$ resolution
and scaled to the main beam temperature scale. All spectra refer
to a 22\arcsec~ beam; the higher resolution $^{12}$CO J=3-2
data have been convolved to this larger beam. The $^{13}$CO J=2-1
spectrum has been scaled up by a factor of 3. \label{fig-1}}

\figcaption[fig2a.ps,fig2b.ps]{(a) A typical large velocity gradient solution
for the average CO line ratios for six giant molecular clouds in
M33. 
The lines indicate the observed $\pm 1\sigma$ values for the line ratios
of: $^{12}$CO/$^{13}$CO J=1-0 ($9.7\pm 0.6$, dashed line);
$^{12}$CO/$^{13}$CO J=2-1 ($7.3\pm 0.5$,
dotted line); $^{12}$CO J=3-2/J=2-1 ($0.69\pm 0.06$, solid line).
Acceptable values of molecular hydrogen density $n_{H_2}$ and $^{12}$CO 
column density per unit velocity $N(CO)/\Delta V$ for a kinetic 
temperature $T_K=20 $ are found
where all three sets of lines intersect.
This figure shows solutions for 
an adopted [$^{12}$CO]/[$^{13}$CO] abundance ratio of 50; slightly warmer
(30 K) solutions are possible if we adopt an abundance ratio of
30. 
(b) A similar large velocity gradient solution
showing the lowest kinetic temperature solution (100 K) for the
cloud NGC 604-2 located in the giant HII region NGC 604. 
The values plotted for the $^{12}$CO/$^{13}$CO J=1-0 and J=2-1 ratios
are the same as in (a); for the $^{12}$CO J=3-2/J=2-1 ratio, the
plotted values are $1.07\pm 0.16$, where the uncertainty includes the absolute
calibration uncertainty.\label{fig-2}}

\figcaption[fig3.ps]{Large velocity gradient solutions with
a kinetic temperature of 15 K for three different assumed values
of the [$^{12}$CO]/[$^{13}$CO] abundance ratio.
This figure demonstrates the existence of  solutions with 
different values of the abundance ratio  but the
same density, kinetic temperature, and $^{13}$CO column density. 
The lines indicate the observed $\pm 1\sigma$ values for the line ratios
of: $^{12}$CO/$^{13}$CO J=1-0 (9.1,10.3, dashed line);
$^{12}$CO/$^{13}$CO J=2-1 (6.8,7.8, 
dotted line); $^{12}$CO J=3-2/J=2-1 (0.63,0.75, solid line).
The vertical solid line indicates the common density solution
of 8300 cm$^{-3}$. (a) [$^{12}$CO]/[$^{13}$CO]=30.
(b) [$^{12}$CO]/[$^{13}$CO]=50. (c) [$^{12}$CO]/[$^{13}$CO]=70.
\label{fig-3}}

\clearpage

\begin{deluxetable}{lccccccc}
\footnotesize
\tablecaption{Log of JCMT Observations \label{tbl-1}}
\tablewidth{0pt}
\tablehead{
\colhead{} & \colhead{} & \colhead{} & \multicolumn{3}{c}
{\underline{$^{12}$CO and $^{13}$CO J=2-1 data}} &  \multicolumn{2}{c}
{\underline{$^{12}$CO J=3-2 data}} \\ 
\colhead{Cloud} & \colhead{$\alpha(1950)$}& \colhead{$\delta(1950)$}& 
\colhead{Date\tablenotemark{a}}   & \colhead{Time}   & 
\colhead{Time}  & \colhead{Date\tablenotemark{b}} & \colhead{Time}  \nl
{} &($^h$ $^m$ $^s$)  & ($^o$ $^\prime$ \arcsec) 
& \colhead{}   & \colhead{($^{12}$CO, $m$)}  
& \colhead{($^{13}$CO, $m$)}  & \colhead{} & 
\colhead{($m$)}  \nl
} 
\startdata
MC 1 & 01:31:03.0 & 30:23:56 & 94/7/17\tablenotemark{c} & 18 
& 72 & 95/8/1 & 60 \nl
MC 13 & 01:31:10.3 & 30:20:40 & 94/7/16-17\tablenotemark{c} 
& 24 & 62 & 95/8/1 & 60 \nl
MC 19 & 01:31:13.4 & 30:23:52 & 93/8/1 & 24 & 60 & 95/8/1 & 60 \nl
MC 20 & 01:31:10.9 & 30:25:24 & 93/8/1 & 24 & 72 & 
95/7/31\tablenotemark{d} & 48 \nl
MC 32 & 01:30:51.8 & 30:23:52 & 93/7/30 & 30 & 42\tablenotemark{e} 
& 95/8/1 & 60 \nl
NGC 604-2 & 01:31:44.2 & 30:31:27 & 95/7/31 & 18 & 48 & 95/7/31 & 30 \nl
NGC 604-4 & 01:31:45.1 & 30:30:58 & 93/7/31 & 18 & 42 & 95/8/1 & 60 \nl
\enddata
\tablenotetext{a}{Observations 1993 July 30 - August 1 obtained with  the
Canadian Acousto-Optical Spectrometer with 0.248262 MHz resolution; other
observations obtained with the Dutch Autocorrelation Spectrometer (DAS) with
0.3125 MHz resolution.}
\tablenotetext{b}{Observations obtained with DAS with 0.078125 MHz
resolution.}
\tablenotetext{c}{Telescope efficiency was 70\% of normal.}
\tablenotetext{d}{Observations obtained with 0.3125 MHz resolution.}
\tablenotetext{e}{Observations obtained on 93/8/26 with 0.15625 MHz resolution.}
\end{deluxetable}

\clearpage

\begin{deluxetable}{lcccccc}
\footnotesize
\tablecaption{CO Line Ratios in Molecular Clouds in M33 \label{tbl-3}}
\tablewidth{0pt}
\tablehead{
\colhead{Cloud} & \colhead{I$(^{12}$CO 2-1)} & \colhead{$\Delta V_{FWHM}$}  
& \colhead{\underline{$^{12}$CO J=2-1}}   & 
\colhead{\underline{$^{12}$CO J=3-2}} & 
\colhead{\underline{$^{12}$CO J=1-0}\tablenotemark{a} }
& \colhead{12+log(O/H)\tablenotemark{b}} 
\nl
\colhead{} & \colhead{(K km s$^{-1}$)}   & \colhead{(km s$^{-1}$)} 
  & \colhead{$^{13}$CO J=2-1}   & \colhead{$^{12}$CO J=2-1} 
& \colhead{$^{13}$CO J=1-0} 
& \colhead{} \nl
} 
\startdata
HII region: & \nl
MC 1  & $6.7\pm0.3$ & 8.0 &  $6.9\pm1.0$ & $0.87\pm0.07$ 
& $9.2\pm2.0$ & 9.02  \nl
MC 13  & $6.8\pm0.4$ & 7.5 & $6.2\pm0.7$ & $0.82\pm0.08$ 
& $10.1\pm0.6$ & 8.83  \nl
MC 20  & $13.6\pm0.2$ & 10.5 & $7.4\pm0.6$ & $0.73\pm0.03$ 
& $9.0\pm0.4$ & 8.91  \nl
NGC 604-2\tablenotemark{c}  & $14.6\pm0.1$ & 11.0 & $9.9\pm0.6$ & $1.07\pm0.03$
& $11.9\pm2.0$  & 8.51  \nl
No HII region: & \nl
MC 19  & $8.2\pm0.4$ & 10.0 & $6.3\pm0.4$ & $0.54\pm0.02$ 
& $11.3\pm0.8$ & 8.98  \nl
MC 32  & $12.2\pm0.3$ & 11.0 & $8.2\pm1.2$ & $0.70\pm0.02$ 
& $8.9\pm0.9$ & 8.83  \nl
NGC 604-4  & $11.8\pm0.4$ & 9.0 & $6.4\pm1.0$ & $0.50\pm0.04$ 
& $7.4\pm0.6$ & 8.51 \nl
\enddata
\tablenotetext{}{Uncertainties are uncertainties in the mean}
\tablenotetext{a}{From \markcite{ww94}Wilson \& Walker 1994}
\tablenotetext{b}{Metallicities interpolated from \markcite{v88}Vilchez 
et al. 1988}
\tablenotetext{c}{This cloud is located in the giant HII region NGC 604}
\end{deluxetable}

\clearpage

\begin{deluxetable}{lcccccc}
\footnotesize
\tablecaption{Large Velocity Gradient Solutions for M33 Clouds
 \label{tbl-4}}
\tablewidth{0pt}
\tablehead{
\colhead{Cloud} & \colhead{[$^{12}$CO]/[$^{13}$CO]} & \colhead{$T_K$}   
& \colhead{$n_{H_2}$}  
& \colhead{$N(^{12}\rm CO)$}   \nl 
\colhead{} & \colhead{}    & \colhead{(K)}   & \colhead{(cm$^{-3}$)} 
 & \colhead{(cm$^{-2}$)} \nl
} 
\startdata
NGC 604-2 & 30 & 200-300 & $2-3\times 10^3$ & $20\times 10^{17}$ \nl
& 50 & 100-300 & $1-3\times 10^3$ & $30-70 \times 10^{17}$ \nl
& 70 & 100-300 & $1-3\times 10^3$ & $50-100\times 10^{17}$ \nl
& & & & \nl
with HII regions & 30 & 30-100 & $2-6\times 10^3$ & $ 5-9 \times 10^{17}$ \nl
(MC 1, MC 13,  & 50 & 20-50 & $2-7\times 10^3$ & $7-20\times 10^{17}$ \nl
MC 20) & 70 & 15-30 & $3-10\times 10^3$ & $7-20\times 10^{17}$ \nl
& & & & \nl
without HII regions & 30 & 10-20 & $5-30\times 10^3$ & $2-4 \times 10^{17}$ \nl
(MC 19, MC 32, & 50 & 10 & $10-20\times 10^3$ & $3\times 10^{17}$ \nl
NGC 604-4) & 70 & 10 & $10\times 10^3$ & $4\times 10^{17}$ \nl
& & & & \nl
average of six clouds 
& 30 & 15-30 & $3-10\times 10^3$ & $2-6\times 10^{17}$ \nl
(without NGC 604-2) & 50 & 10-20 & $4-30\times 10^3$ & $3-6\times 10^{17}$ \nl
& 70 & 10-20 & $6-20\times 10^3$ & $4-6\times 10^{17}$ \nl
\enddata
\end{deluxetable}

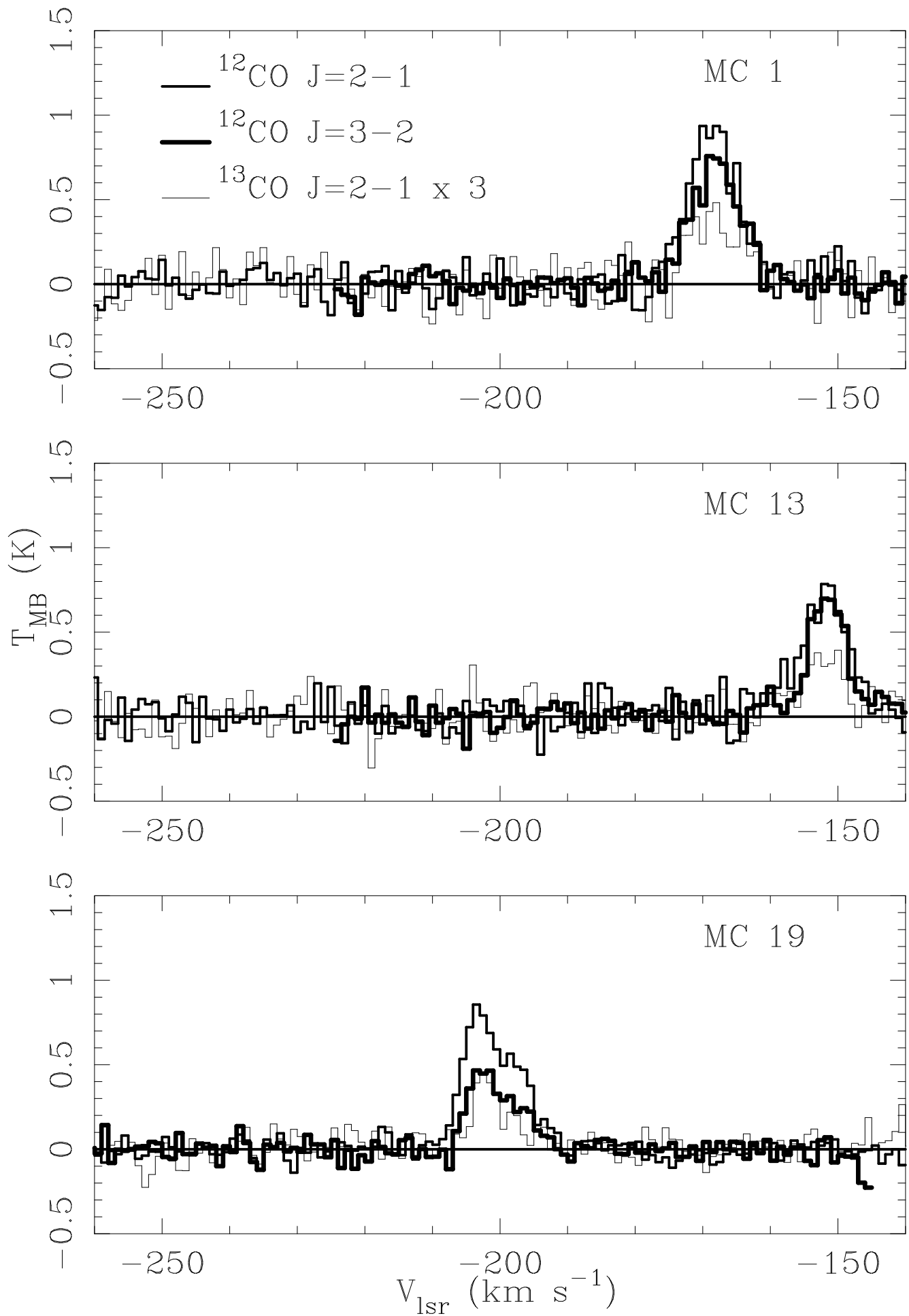
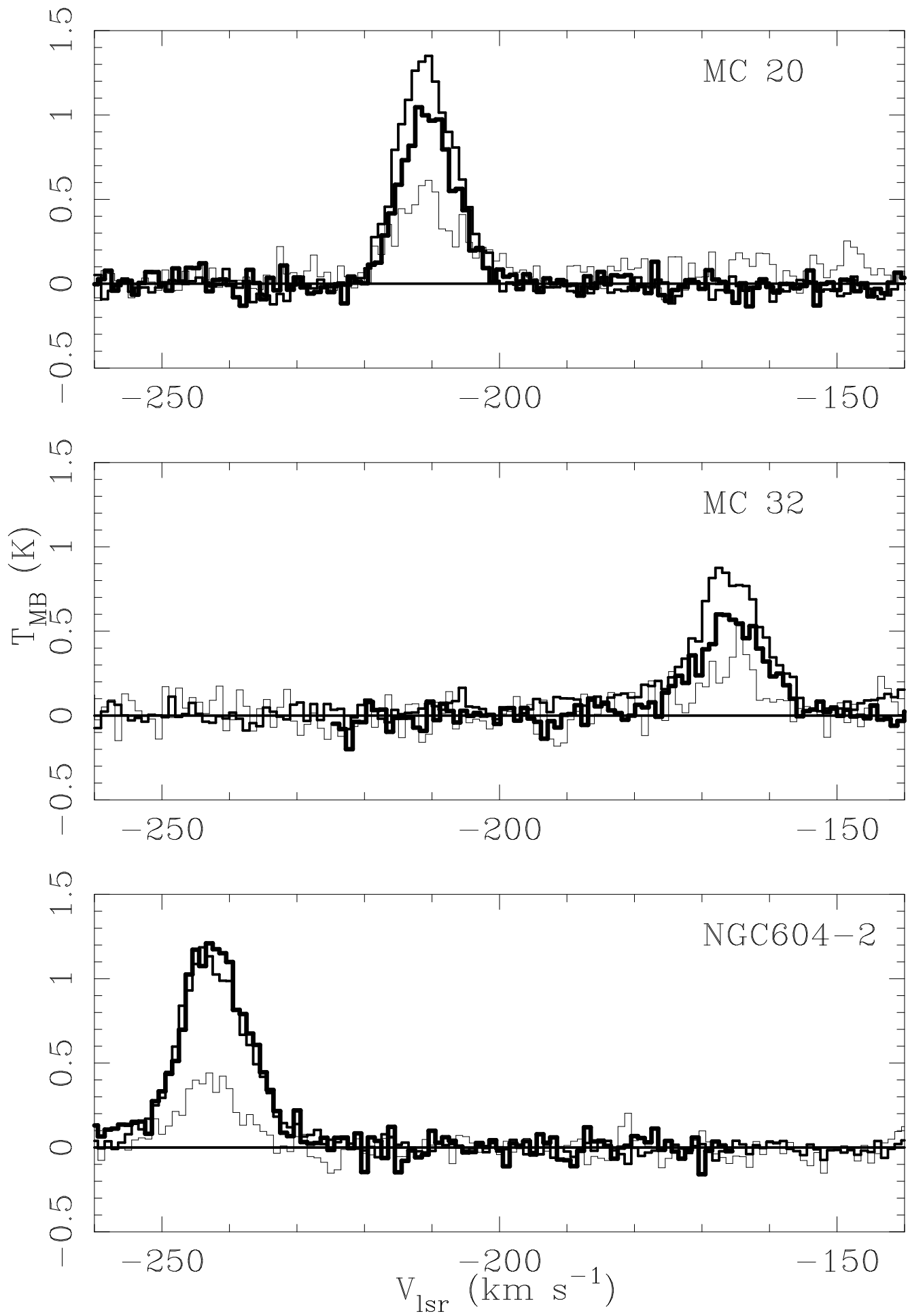
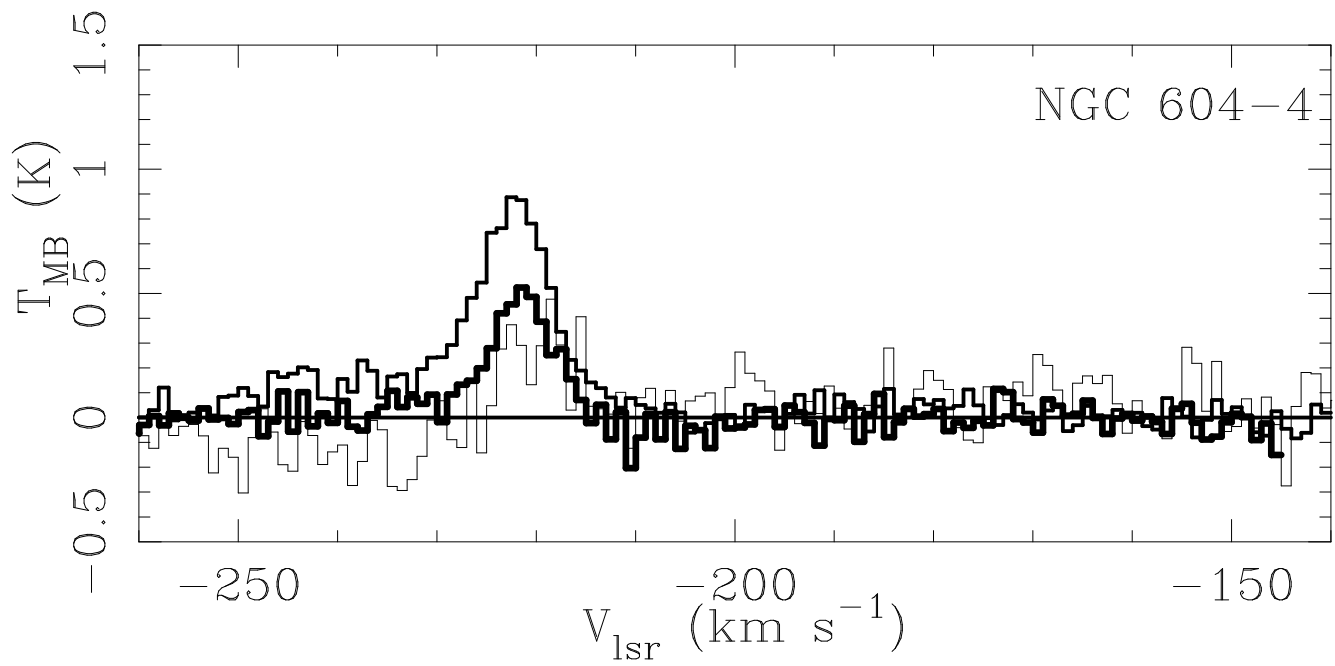
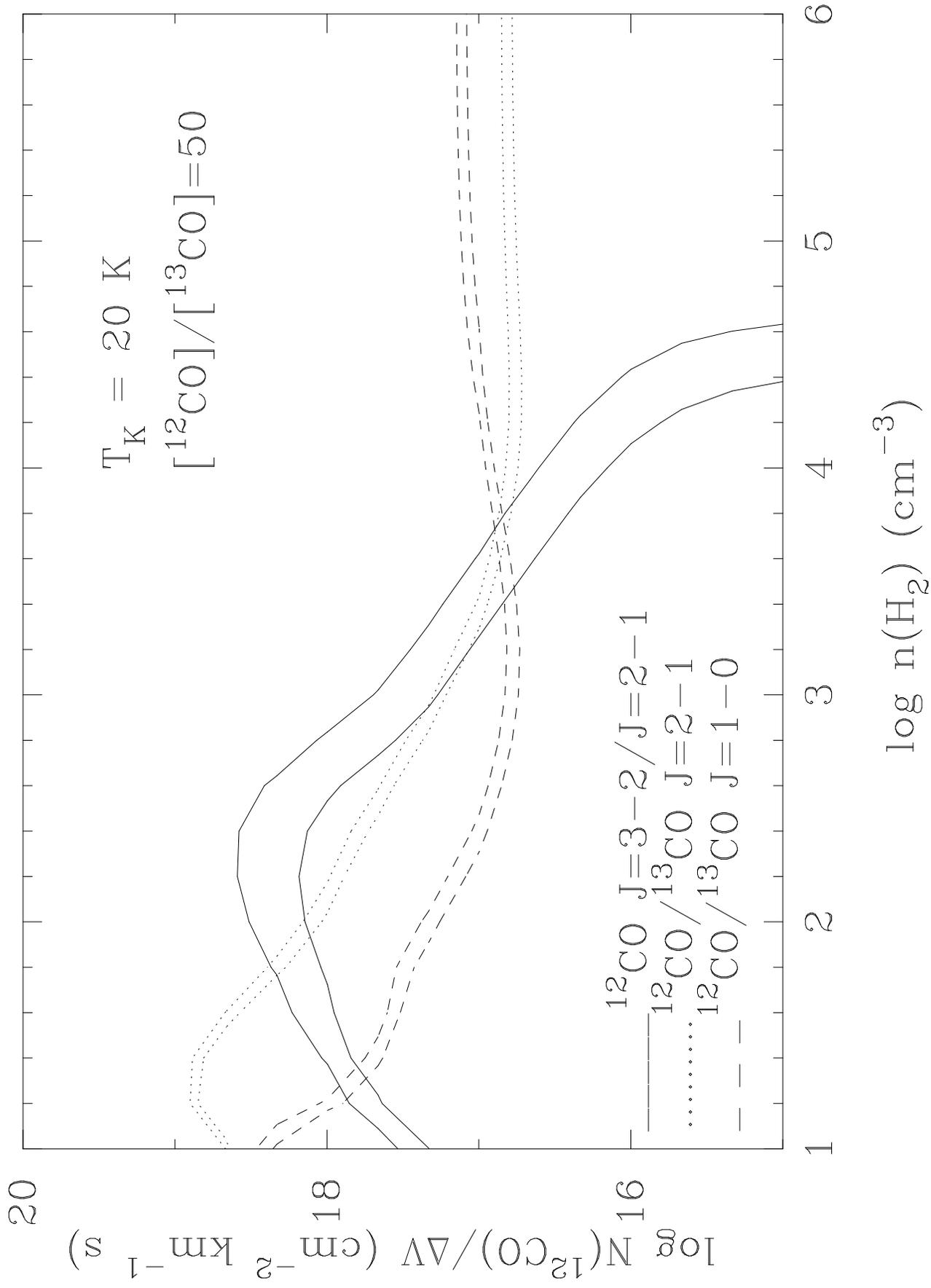
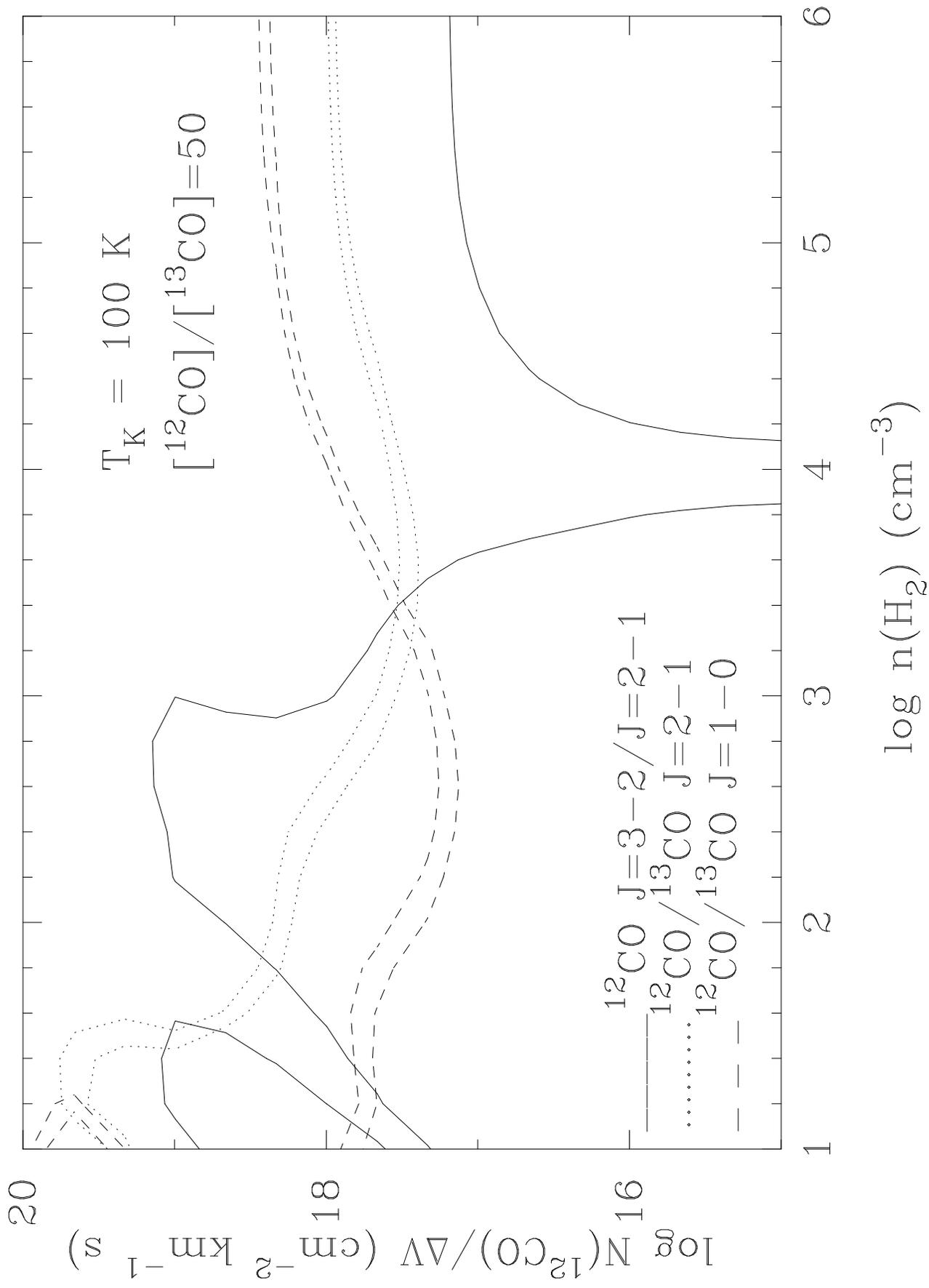
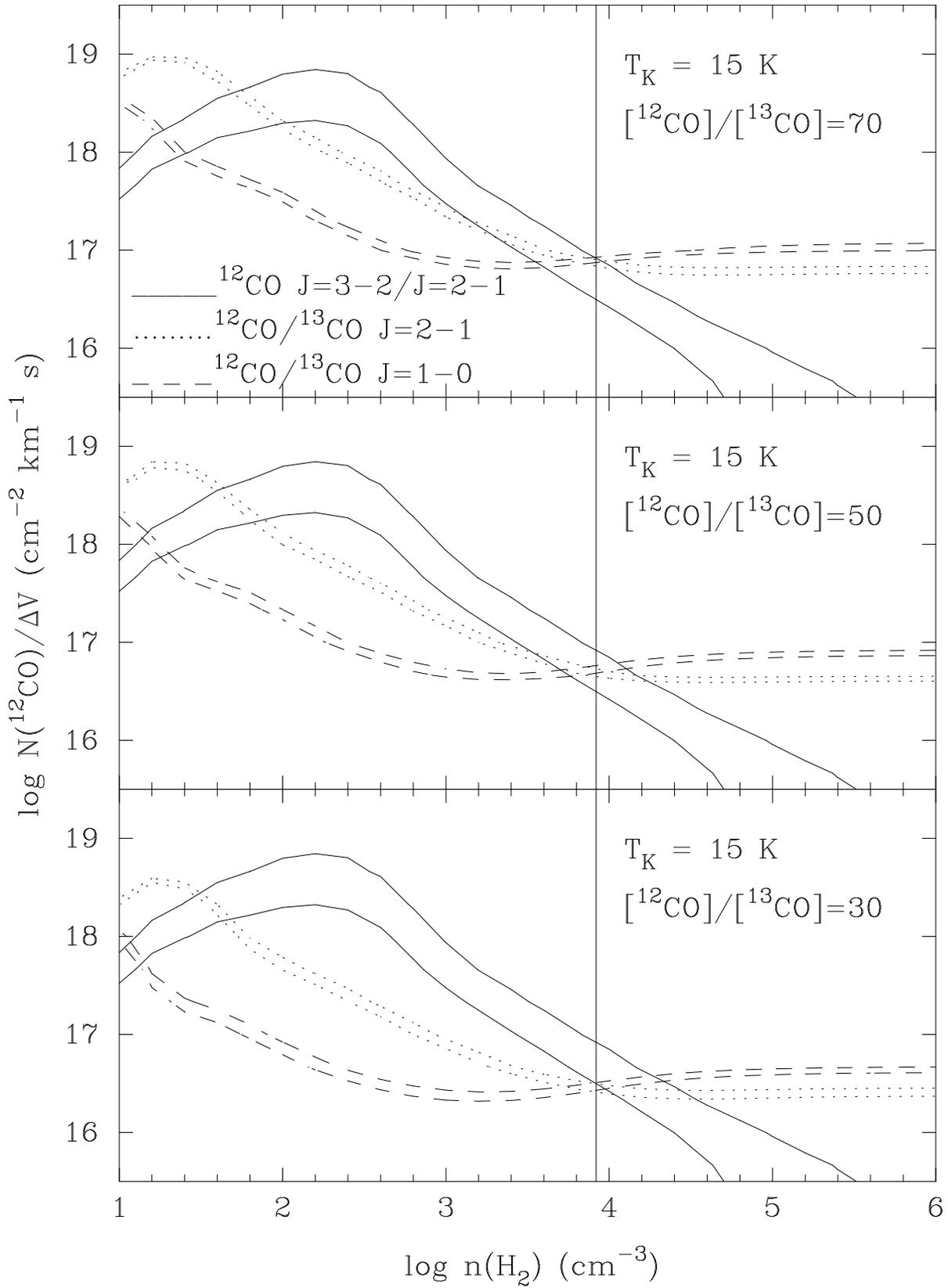
\end{document}